\documentclass[floatfix,useAMS,usenatbib]{mn2e}
\usepackage{graphicx,epsfig}
\usepackage{multirow}
\usepackage{verbatim}
\usepackage[figuresright]{rotating}

\usepackage[T1]{fontenc}
\usepackage{aecompl}

\usepackage{graphicx}
\usepackage{graphpap}
\usepackage{epsf}
\usepackage{amssymb}
\usepackage{color}

\def\FIG #1 #2 [#3] #4\par{%
  \begin{figure}\begin{center}%
    \includegraphics*[#3]{#2}%
    \\
    \caption{#4}%
    \label{#1}%
  \end{center}\end{figure}%
}
\def\FIGG #1 #2 #3 [#4] #5\par{%
  \begin{figure*}[ht]\begin{center}%
               \includegraphics*[#4]{#2}
               \includegraphics*[#4]{#3}
    \caption{#5}%
    \label{#1}%
   \end{center}\end{figure*}%
}
\def\FIGth #1 #2 #3 #4 [#5] #6\par{%
  \begin{figure*}[ht]\begin{center}%
        \includegraphics[#5]{#2} 
        \includegraphics[#5]{#3}
        \includegraphics[#5]{#4}
        \caption{#6}
        \label{#1}
   \end{center}\end{figure*}
}
\def\FIGfo #1 #2 #3 #4 #5 [#6] #7\par{%
  \begin{figure*}[ht]\begin{center}%
        \includegraphics[#6]{#2}
        \includegraphics[#6]{#3} \\
        \includegraphics[#6]{#4}
        \includegraphics[#6]{#5}
        \caption{#7}
        \label{#1}
   \end{center}\end{figure*}
}
\def\FIGfi #1 #2 #3 #4 #5 #6 [#7] #8\par{%
  \begin{figure*}[ht]\begin{center}%
        \includegraphics[#7]{#2}
        \includegraphics[#7]{#3} 
        \includegraphics[#7]{#4} \\
        \includegraphics[#7]{#5}
        \includegraphics[#7]{#6}
        \caption{#8}
        \label{#1}
   \end{center}\end{figure*}
}
\def\FIGsi #1 #2 #3 #4 #5 #6 #7 [#8] #9\par{%
  \begin{figure*}[ht]\begin{center}%
        \includegraphics[#8]{#2}
        \includegraphics[#8]{#3}\\
        \includegraphics[#8]{#4}
        \includegraphics[#8]{#5}\\
        \includegraphics[#8]{#6}
        \includegraphics[#8]{#7}
        \caption{#9}
        \label{#1}
    \end{center}\end{figure*}
}

\def\rfig#1{Figure \ref{#1}}

\def\msun{M_\odot}

\newcommand{\mt}[1]{\mathrm{#1}}
\newcommand{\sect}[1]{Section~\ref{#1}}
 
\def\lvm{\leavevmode\hbox to\parindent{\hfill}}

\def\BE{\begin{equation}}
\def\EE{\end{equation}}
\def\BA{\begin{array}}
\def\EA{\end{array}}
\def\BAN{\begin{eqnarray*}}
\def\EAN{\end{eqnarray*}}

\def\la{\mathrel{\mathpalette\fun <}}
\def\ga{\mathrel{\mathpalette\fun >}}
\def\fun#1#2{\lower3.6pt\vbox{\baselineskip0pt\lineskip.9pt
\ialign{$\mathsurround=0pt#1\hfil##\hfil$\crcr#2\crcr\sim\crcr}}}

\def\ergs{ergs s$^{-1}$ }

\def\kms{km s$^{-1}$}
\def\msun{M_\odot}

\def\e#1{$\times 10^{#1}$ }
\def\ee#1{$10^{#1}$ }

\def\ltsima{$\; \buildrel < \over \sim \;$}
\def\ltsim{\lower.5ex\hbox{\ltsima}}
\def\gtsima{$\; \buildrel > \over \sim \;$}
\def\gtsim{\lower.5ex\hbox{\gtsima}}

\def\nism{n_0}
\def\bism{B_0}
\def\cc{\,\mt{cm}^{-3}}
\def\geff{\gamma_\mt{eff}}

\def\ufs{u_0}
\def\pinj{p_\mt{inj}}
\def\pmax{p_\mt{max}}
\def\rtot{r_\mt{tot}}

\def\Eth{E_\mt{th}}
\def\Ekin{E_\mt{kin}}
\def\Ecr{E_\mt{CR}}
\def\Pcr{P_\mt{CR}}

\def\tcr{t_\mt{CR}}

\def\aap{A\&A}
\def\apjs{ApJ Supl.}

\usepackage{tikz}

\newcommand\basedir{graphs}

\usepackage{txfonts}
\usepackage{natbib}
\bibpunct{(}{)}{,}{a}{}{,}

\def\wfourplotfirst{0.278}
\def\wfourplotfirstlab{0.245}
\def\wfourplotrest{0.217}

\def\marginfirst{0}
\def\margintrest{117}
\def\marginfirstlab{60}

\def\wthreeplotfirst{0.3507}
\def\wthreeplotrest{0.28}

\def\marginslice{40}
\def\wtwoplotfirst{0.455}
\def\wtwoplotrest{0.42}

\begin{document}

\title[Parametric studies of SNRs]{Parametric studies of cosmic ray acceleration in supernova remnants}

\author[D. Kosenko,  G. Ferrand, A. Decourchelle]{D. Kosenko$^{1}$\thanks{E-mail:
daria.kosenko@gmail.com}, G. Ferrand$^{2}$, A. Decourchelle$^{3}$\\ 
$^{1}$Sternberg Astronomical Institute, M.V.Lomonosov Moscow State University, Universitetskij pr. 13, 119992 Moscow, Russia \\
$^{2}$Department of Physics \& Astronomy, University of Manitoba, Winnipeg, MB, R3T 2N2, Canada, \\
$^{3}$Laboratoire AIM (CEA/Irfu, CNRS/INSU, Universit\'e Paris VII), CEA Saclay, b\^at. 709,  F-91191, Gif sur Yvette, Cedex, France\\
}

\maketitle

\begin{abstract}
We present a library of numerical models of cosmic-ray accelerating supernova remnants (SNRs) evolving through a homogeneous ambient medium. We analyze distributions of the different energy components and diffusive shock acceleration timescales for the models in various conditions.
The library comprises a variety of SNR evolutionary scenarios and is used to map remnants with sufficiently known properties. This mapping constrains the respective ambient medium properties and the acceleration efficiency. Employing the library, we derive the ambient medium density, ambient magnetic field strength and the cosmic-ray acceleration efficiency for models of Tycho and SN~1006 remnants and refine the ages of SNR~0509-67.5 and SNR~0519-69.0. 
\end{abstract}


\section{Introduction}

Supernova remnants (SNRs) are believed to be the sites of particles acceleration and main contributors to the observed cosmic rays (CR) spectrum up to energies of \ee{15.5}eV \citep[e.g.][]{arnett73,blandford87,berezhko07}. However, a direct observational evidence showing that they can accelerate protons up to the CR spectrum ``knee'' was found only in the remnant of SN~1752 (G120.1+1.4, observed by Tycho Brahe, henceforth Tycho). The data from the Chandra X-ray observatory revealed the presence of \ee{14}$-$\ee{15}eV protons \citep{eriksen11}. The first evidence of high-energy electrons accelerated by the forward shock (FS) of the remnant of SN~1006 (G329.6+14.6) was found by \citet{koyama95} in analysis of the non-thermal X-ray emission observed with ASCA. The detailed multi-wavelength study of \citet{cassam-chenai08} confirmed the acceleration of ions in this SNR.

Significant progress has been made in the last decades in developing the theory of nonlinear diffusive shock acceleration (NLDSA), allowing to accurately account for the CR feedback on the SNR evolution \citep{drury83,blandford87,jones91,malkov01}. In spite of the substantial development in the detailed small-scale Fermi acceleration modeling \citep{bell13} the self-consistent simulations of this process in SNRs still require parametrization \citep[e.g. see the reviews of][and the references therein]{schure12, helder12}. 

The efficiency of the NLDSA acting in SNRs is affected by the ambient medium conditions such as density and magnetic field. Moreover, the properties of the CR spectrum strongly depend on the evolutionary stage of the remnant. On the other hand, the back-reaction of the accelerated CRs on the shock structure plays an important role in the SNR's evolution, as a non-negligible part of the supernova kinetic energy escapes from the remnant in the form of the relativistic particles. To estimate the effect of this back-reaction on the SNR evolution in different conditions, a detailed modeling has to be performed. Thus, we created a library of numerical evolutionary models for SNRs based on a grid of the ambient density values, a parameter regulating the CR acceleration efficiency, and for different ambient magnetic field strengths. This library allows to study the evolution of the NLDSA properties in the models under different conditions. 

Multi-wavelength observations of young SNRs and theoretical evolutionary scenarios suggest that the remnants of type~II and Ib/c supernovae, results of the core collapse of massive stars, evolve through a non-homogeneous circumstellar medium modified by an outflow from the supernova progenitor. Type Ia supernovae, results of the thermonuclear explosions of white dwarfs, are usually located in a rather uniform medium, largely unaffected by the progenitor wind.
In spite of general observational resemblance, young thermonuclear SNRs at the ejecta-dominated or Sedov stages \citep{chevalier82} evolving through a relatively homogeneous medium expose a substantial diversity of their morphologies and emission properties. The reason can be either different evolutionary stages of the remnants together with the different ambient conditions or, possibly, different explosion scenarios. It is often difficult to distinguish between the morphological irregularities caused by the supernova origin or by its interaction with the turbulent medium \citep[][]{lopez13}. Hence, in our simulations we fix the explosion mechanism ($E_0=10^{51}\,\mt{erg}$, power-law ejecta density distribution with index $n=7$) and study only the effects of the interaction with the ISM.

This set of SNRs evolutionary  scenarios, covering a wide ranges of the ISM densities, magnetic fields, and diffusive shock acceleration (DSA) efficiencies can be used to constrain an allowed parameter space for an SNR with known properties. We considered two well-studied nearby type Ia SNRs, which are Tycho and SN~1006. We used the measured dynamical properties such as FS radius and expansion velocity to constrain the ambient medium density, and properties of the SNR shells to constrain the efficiency of the CR acceleration. We considered as well two SNRs in the Large Magellanic Cloud (LMC): SNR~0509-67.5 and SNR~0519-69.0. The FS radii and velocities were used to constrain the ambient medium densities and the ages for each of these remnants.

The structure of the paper is as follows. In \sect{sec:method} we describe our numerical code and the method. In \sect{sec:grid} we present the grid of the initial parameters. We report the energy evolution profiles and timescales in \sect{sec:ecr}. \sect{sec:galactic} contains the analysis of the Galactic SNRs and \sect{sec:lmc} presents models for the remnants in the LMC.  We outline the results and discuss them in \sect{sec:results} and we conclude by \sect{sec:conclusion}.

\section{Description of the method}
\label{sec:method}
The simulations of SNRs evolution are performed with the multidimensional hydrodynamical code {\sc ramses} \citep{teyssier02} taking advantage of the adaptive mesh refinement (AMR) technique. The method was developed by \citet{ferrand10} where they coupled the detailed calculation of the NLDSA through a semi-analytic kinetic approach \citep[introduced by][]{blasi02} with the {\sc ramses} hydrodynamical solver. 

Technical realization of the method is as follows \citep{ferrand10}. The NLDSA routine is executed at every time-step of the hydrodynamical simulation. Given the shock properties (velocity and upstream conditions) provided by the {\sc ramses} hydrodynamical solver, Blasi's model jointly solves the particle spectrum and the fluid velocity profile. 
We recall that Blasi's model essentially connects any point in the upstream with the downstream, for both the relativistic particles and the fluid, and thus predicts the compression ratios \citep[see][]{blasi02, blasi05}. At the sub-shock, it uses the conservation of momentum, including the pressure of CRs and of magnetic field (MF) waves, where the general formula for the pressure jump is given by equation (10) of \citet{caprioli09}. To get the overall compression, a hypothesis on the behaviour of the fluid in the precursor is necessary. The most common one is the Alfv\'en heating. It depends on the value of the parameter $\zeta$, which was introduced in \citet[][equation 47]{caprioli09}. $\zeta=0$ implies effective magnetic field amplification (MFA), and $\zeta=1$ corresponds to the case where amplification of the magnetic field is severely damped due to the transfer of the magnetic energy in the form of waves into thermal energy of the plasma.\footnote{Note that the turbulent heating in the case of non-linear Landau damping is ineffective in the fast shocks expanding into not very hot medium, and relevant only for the remnants expanding into pre-supernova hot winds or in old SNRs \citep{caprioli09,volk81}.} 
The general formula that links $r_\mt{sub}$ and $\rtot$ is given by equation (50) of \citet{caprioli09}.

Thus with the Mach number $M_\mt{S}$ provided by the hydro code and the total compression ratio $\rtot$ from the Blasi model, using the equation (6) of \citet{ellison04} we calculate the effective adiabatic index at the shock $\geff$, that is applied upstream of the shock front. Then the hydro solver naturally generates the shock predicted by the acceleration model, by integrating the usual conservation equations. 

Note that it can only find stationary solutions, that we update after each time-step, given the current flow conditions and the maximum energy particles have reached at that time. 
The effective adiabatic index is applied to the gas equations in the {\sc ramses} hydrodynamical scheme just upstream of the FS, in such a way that each fluid element retains its $\geff$, advecting it inside the shocked region (the diffusion of the relativistic particles is not accounted for downstream of the shock). The NLDSA model is activated only at the FS of the remnant. 

%

Injection of the relativistic particles from the downstream thermal pool is regulated by parameter $\xi=p_\mt{inj}/p_\mt{th}$ \citep{blasi05}, where $p_\mt{th}$ is the downstream thermal momentum of the Maxwellian distribution, and $\pinj$ is the injection momentum, so that all particles with $p>\pinj$ contribute to the acceleration process. Typically $\xi$ can be in the range $3.0-4.5,$ thereby high values of $\xi\gtrsim4.0$ correspond to the test-particle regime with weak back-reaction of the accelerated particles on the shock dynamics and low values of $\xi\lesssim3.5$ imply efficient DSA. 
The relation between the injection parameter $\xi$ and the fraction of the injected relativistic particles $\eta$ is given by \citep{blasi05} 
\begin{equation}
\eta = \frac{4}{3\pi^{1/2}} \left(R_\mt{sub} - 1\right)\xi^3\,e^{-\xi^2}
\label{xieta}
\end{equation} 
where $R_\mt{sub}$ is the compression ratio at the subshock.

In addition, 
we consider the effects of the efficiency of Alfv\'en waves heating in the precursor on the dynamics, $\zeta=0$ or $1$. 
Although the Alfv\'enic drift of the upstream CR scattering centers is parametrized in the code employing the recipe of \citet{kang13}, we do not investigate this effect in the current study. 

\section{Parametric grid}
\label{sec:grid}
In the grid for the SNR evolutionary models the ambient density $\nism$ ranges from 0.01 to 7 $\cc$. This range covers completely the typical values of the ambient densities, measured in the vicinity of type Ia SNRs (e.g. with $\sim0.05\cc$ for SN~1006 and $\sim3\cc$ for the 0519-69.0 SNRs). The CR injection momentum parameter $\xi$ runs from 3.1 to 4.1, spanning the typical range of values $\xi = 3.5-3.9$ inferred for observed SNRs. We consider configurations with the ambient magnetic field $\bism = (0.3 - 30)\,\mu$G, which are typical for the SNR ambient medium. By default we set Alfv\'en wave damping efficiency $\zeta=0$ in most of the simulations, however a few models with $\zeta = 1$ were analyzed additionally.

The analytical solution from \citet{chevalier82, chevalier83}  
is used as the initial conditions for the runs. 
%
This condition is relevant for the remnants of type Ia supernova, evolving through the uniform ambient medium. Normally, we start the simulations at the age of 30 years. The explosion energy $E_0$ is set to \ee{51}erg, the mass of the ejecta of $1.4\,\msun$. We consider a warm ($T_0=$\ee{4}K) hydrogen medium with the mean particle weight of the proton mass.
The maximal resolution of the AMR numerical mesh is of $128^3$ cells. It corresponds to the minimal cell size of 0.04, 0.08, 0.16 pc for a remnant's age of 200, 800, 2600 yr.
For the detailed modeling of Tycho and SN~1006 we performed a few runs of higher spacial resolutions with minimal cell sizes of $0.02$ and $0.05$ pc correspondingly. The library comprises more than 300 hydrodynamical models.

%
%

\def\zetaval{00}
\def\zeval{ $ \makehalf{\zetaval}\mymathresult $ }
\def\lmcsnr{}
\def\crfl{}

\section{Evolution of the DSA parameters}
\label{sec:ecr}
The evolution of the relevant DSA parameters for $\nism=0.03,0.3\cc$  and $\bism=1,\,10\,\mu$G is presented in \rfig{fig:crpar_gal_\zetaval_}. From top to bottom the quantities are as follows. The CR pressure immediately upstream of the shock normalized  to the total pressure $\Pcr/P_\mt{tot}$,  the proton injection momentum $\pinj$ (in units of $m_pc$, $m_p$ proton mass, $c$ speed of light), the  proton maximum momentum $\pmax$ (in units of $m_pc$), the compression ratio at the subshock $r_\mt{sub}$, the total compression ratio $r_\mt{tot}$.  
The colours correspond to different values of~$\xi$. Models with $\xi=4.1, 3.9, 3.7, 3.5, 3.1$ are plotted with green, yellow, red, magenta and blue lines correspondingly. Solid lines are for $\zeta=0$, dashed lines correspond to $\zeta=1$.
The plots with acceleration efficiency $\Ecr/\Ekin$ and the SNR energy evolution (total -- thick solid line, kinetic -- dotted, thermal -- dashed, CR -- thin solid) are presented in the two bottom rows, where only the models with $\zeta=0$ are shown, and the cases for the $\xi=4.1, 3.7, 3.1$ are outlined with green, red and blue lines respectively.  Note that the evolution of the total energy (thick solid line) deviates from the expected constant value of \ee{51} erg, due to effects of the numerical integration over the volume of the low-resolution models. These profiles are shown for comparison of the cases with various $\xi$.

According to \citet{be99} 
a fast shock becomes modified by the CRs when the following condition is met
\begin{equation}
\frac{2\sqrt{10}}{3}\,\eta\, \rtot\, \pinj\, \pmax^{1/4} \left(\frac{c}{\ufs}\right)^2 \simeq 1
\label{be99_37}
\end{equation} 
where we assume a non-modified case of $r_\mt{sub}=4$, all momenta are in units of $m_pc$,  and $\ufs$ is the FS speed (in \kms). This expression yields the corresponding  value for the injection momentum as 
\begin{equation}
\pinj = 10^{-3} (\ufs/10^3)^2\,(\eta/10^{-4})^{-1}\,(\rtot/4)^{-1}\,(\pmax/10^4)^{-1/4}.
\label{pinjcrit}
\end{equation} 
We recall that the injection momentum at the shock is defined as $\pinj = \xi p_\mt{th}$, where $p_\mt{th}$ is thermal momentum of the plasma downstream of the FS. The shock starts to be CR-modified as soon as the injection momentum reaches $\pinj$. Hence, taking into account that $p_\mt{th} = \sqrt{2 m_p kT}$ ($k$ Boltzman constant, $T$ downstream proton temperature) and using for a non-modified shock $kT = 3/16\,m_p\ufs^2$ we conclude that the transition from test-particle regime to the NLDSA occurs when
\begin{equation}
\ufs \lesssim 2\times10^3\, (\eta/10^{-4})\,(\rtot/4)\,(\pmax/10^4)^{1/4}
\label{ufscrit}
\end{equation} 


The evolutionary plots in the top row of \rfig{fig:crpar_gal_\zetaval_} show that drastic increase of $\Pcr/P_\mt{tot}$ takes place earlier for the models with denser ambient medium and lower values of $\xi$-parameter. According to the contour-plot in the right-hand side of \rfig{fig:tmax_\zetaval} the values of  $\xi=4.1, 3.7, 3.1$ correspond to $\eta/10^{-4} = 0.2, 2, 20$ and consequently the respective critical velocities are $400, 4000, 40000$~\kms. The first two cases can represent remnants of $10^{4},\, 500$ yr expanding through a $0.3\cc$ medium.  
The irregular behaviour of  $\Pcr$ can be due to uncertainties in estimating the FS location, which partially can be caused by numerical instabilities at the shock front (the presented profiles are already smoothed using a median filter). Note that this issue appears when the CR pressure starts to increase rapidly (top left plots of \rfig{fig:crpar_gal_\zetaval_}), leading to strong shock modifications.

Comparison of the models with different ambient magnetic field reveals that the CR pressure is lower for the higher $\bism$, where the amplification is more efficient and a considerable amount of energy is transferred to the amplified magnetic field.
According to \citet{be99} $\rtot = 1.5M_\mt{A0}^{3/8}$ which yields $\rtot \propto (\ufs/\bism)^{3/8}\nism^{3/18}$. This dependence of the compression ratio on the FS velocity, ambient magnetic field and ambient density is reflected in the behaviors of $\Pcr/P_\mt{tot} $. A more detailed investigation of how $\bism$ influences the NLDSA is reported in \citet{ellison05}.
We note that the models with $\zeta=1$ with the efficient wave damping only slightly differ from the models with $\zeta=0$. The effect is visible only for the extremely high efficiencies of $\xi<3.7$ and reduces $\Pcr$ and $\rtot$ only by a few percent. Similar effects of the wave heating on the shock properties were reported by \citet{kang10}.

For the over-magnetized tenuous plasma $r_\mt{tot}\propto (\ufs/\bism)^{3/8}\nism^{3/18}$ is small and leads to rather high $\pinj$. Hence, when $\ufs$ drops considerably (at the age of a few thousands of years), the NLDSA mechanism switches off and results in the transition of the CR modified shock back to the un-modified regime. This is reflected in the energy evolution plots for the $\nism=0.03\cc$, $\bism=10\,\mu$G models (lower plots in the third column of  \rfig{fig:crpar_gal_\zetaval_}). Moreover, for the case of $\nism=0.3\cc$ and $\bism=1\,\mu$G the situation with $r_\mt{tot}$ and $\pinj$ is inverse, this is why the respective acceleration efficiency increases even after $10^4$ yrs (second column of \rfig{fig:crpar_gal_\zetaval_}).

As was noted in \citet{berezhko04} the efficiency of the acceleration gradually declines with time with the decrease of the Alfv\'en Mach number. They have pointed out that for $t>3\times 10^4$~yr the CR production does not play an important role in the SNR dynamics. To improve on this estimate we calculated the timescales when the CR pressure upstream of the FS reaches its maximum. The corresponding maps on the ($\nism$, $\xi$)-grid are presented in the left panel of \rfig{fig:tmax_\zetaval}. The middle panel shows the $t_\mt{CR}$ values normalized by the Sedov timescale $t_\mt{s} = 248.0\, \nism^{-1/3}$~yr 
and the right panel shows the maps of maximum injection parameter. E.g. for $\xi=3.7$, the age when DSA efficiency is still high but starts to fall off is of the order of a few thousands of years and is about $(0.01-0.1)$ of the radiative timescale.


\newcommand\evolplot[6]{
        \includegraphics*[trim=#1 #2 50 #3,clip=True,width=#4\hsize]{\basedir/evol/z\zetaval_b#5/n#6_xi41/\evolfile}%
}

\def\addlab{}
\newcommand\revolplot[2]{
	\evolplot{\marginfirst}{#1}{#2}{\wfourplotfirst}{001}{003}%
	\evolplot{\margintrest}{#1}{#2}{\wfourplotrest}{001}{030}~%
	\evolplot{\margintrest}{#1}{#2}{\wfourplotrest}{010}{003}%
	\evolplot{\margintrest}{#1}{#2}{\wfourplotrest}{010}{030}\\
}
\if ==



 \begin{figure*}
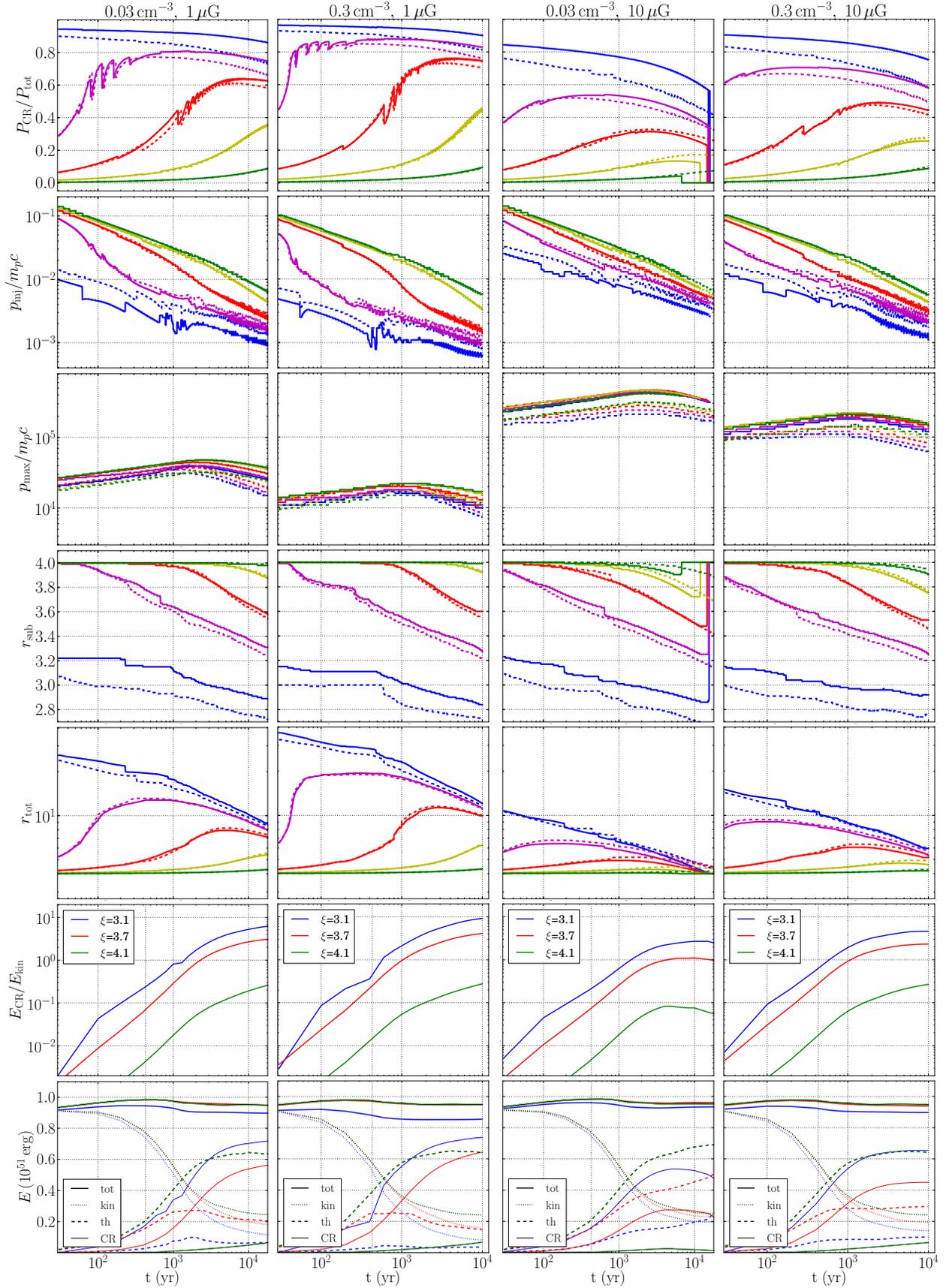
\begin{center}%
 \def\evolfile{Pc_Ptot_med15_tyr}
	\revolplot{62}{1}
\def\evolfile{p_inj_med15_tyr}
	\revolplot{62}{40}
\def\evolfile{p_max_med15_tyr}
	\revolplot{62}{40}
\def\evolfile{r_sub_med301_tyr}
	\revolplot{62}{40}
\def\evolfile{r_tot_med301_tyr}
	\revolplot{62}{40}
\def\evolfile{tr_e_ecr_eng_3D_cell_size7}
	\revolplot{62}{40}
\def\evolfile{tr_e_sum_eng_3D_cell_size7}
	\revolplot{0}{40}
        \caption{Evolution of the DSA properties. Green, yellow, red, magenta and blue lines correspond to $\xi=4.1, 3.9, 3.7, 3.5, 3.1$ models. Solid lines are for $\zeta=0$, dashed lines correspond to $\zeta=1$. The bottom rows contain plots with evolution of the different types of energy ($\zeta=0$):  CR (dash-dotted),  thermal (dashed), kinetic (dotted), and total SNR energy (solid). 
        }
        \label{fig:crpar_gal_\zetaval_\addlab}
    \end{center}\end{figure*}

\else

\newcommand\revolplot[4]{
 \def\evolfile{Pc_Ptot_med9_tyr}
	\evolplot{1}{#1}{#2}{\wthreeplotrest}{#3}{#4}%
\def\evolfile{tr_e_ecr_size7}
	\evolplot{1}{#1}{#2}{\wthreeplotrest}{#3}{#4}%
\def\evolfile{tr_e_sum_size7}
	\evolplot{1}{#1}{#2}{\wthreeplotrest}{#3}{#4}\\
}

\def\evolfile{tr_e_sum_size8}

 \begin{figure*}[ht]\begin{center}%
	\revolplot{32}{1}{010}{030}
	\revolplot{32}{1}{001}{003}
	\revolplot{1}{1}{010}{003}
        \caption{Evolution of the energy and the pressure. The profiles are presented for the various acceleration efficiencies with $\xi=3.1$ (blue), $\xi=3.7$ (red) and $\xi=4.1$ (black).  In the top and the middle row the solid lines denote the models with $\zeta=0$, the dashed lines correspond to the cases with $\zeta=1$..}
        \label{fig:ecr_gal_\zetaval}
    \end{center}\end{figure*}

\fi

%


To estimate the amount of energy expelled into the ISM in our SNR models we calculated the kinetic, thermal, and trapped CR energy components at the age of a remnant a few of $t_\mt{CR}$ (end of the Sedov-Taylor stage). The corresponding maps are presented in \rfig{fig:pcrmax_\zetaval}.
The respective energy shares for the models with $3.6 < \xi <3.8$ are summarized in table~\ref{tab:ecr}.

\newcommand\pcplot[5]{
        \includegraphics*[trim=#1 #2 50 0,clip=true,width=#3\hsize]{\basedir/evol/z00_b#4/#5_max_denxi}%
}

\newcommand\bpcplot[2]{
	\pcplot{\marginfirst}{#1}{\wfourplotfirst}{#2}{time}%
	\pcplot{\margintrest}{#1}{\wfourplotrest}{#2}{stime}~~ 
	\pcplot{\margintrest}{#1}{\wfourplotrest}{#2}{eta_max} \\
}
  \begin{figure*}
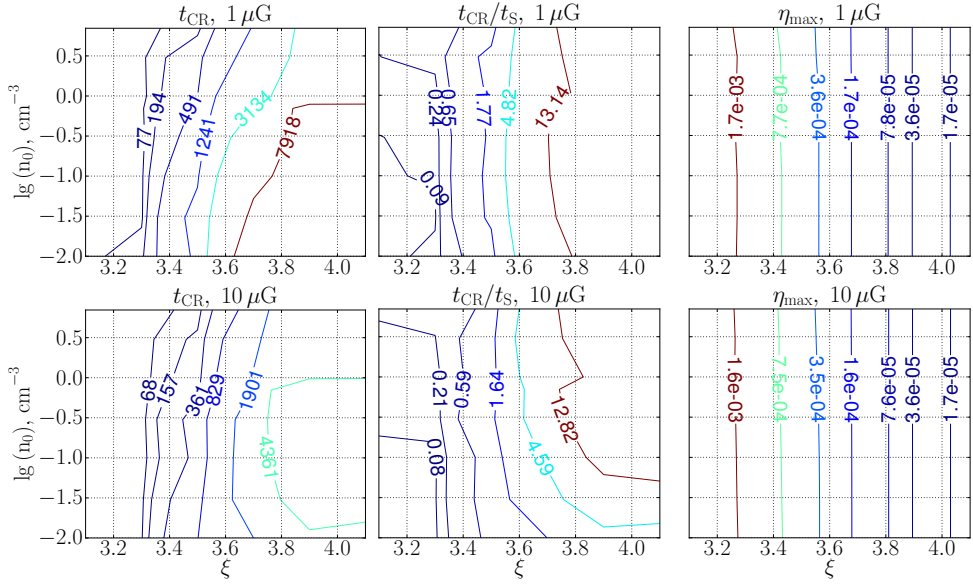
\begin{center}%
        \bpcplot{37}{001}
        \bpcplot{5}{010}
        \caption{Contour maps for the SNR ages when $\Pcr$ reaches its maximum. The value are presented in years (left column) and in Sedov times (middle column). The right column contains maximum values of injection parameter $\eta$. 
        }
        \label{fig:tmax_\zetaval}
    \end{center}\end{figure*}

\renewcommand\bpcplot[2]{
	\pcplot{\marginfirst}{#1}{\wfourplotfirst}{#2}{E_kin}%
	\pcplot{\margintrest}{#1}{\wfourplotrest}{#2}{E_th}%
	\pcplot{\margintrest}{#1}{\wfourplotrest}{#2}{E_CR}%
	\pcplot{\margintrest}{#1}{\wfourplotrest}{#2}{E_lost}\\%
}
  \begin{figure*}
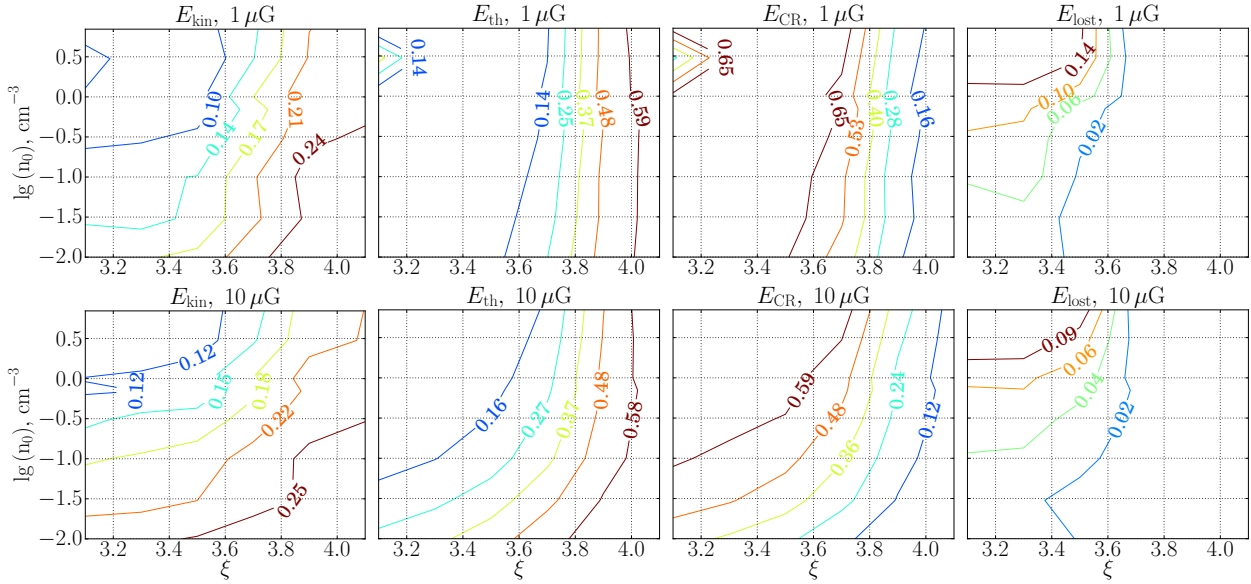
\begin{center}%
        \bpcplot{37}{001}
        \bpcplot{5}{010}
        \caption{Contour maps of the current kinetic energy $\Ekin$ (first column), thermal energy $\Eth$ (second column), total amount of $\Ecr$ that a SNR can generate during its lifespan (third column), and the energy lost from the system due to the CR escape $E_\mt{lost}$(the last column). All the values are in units of the explosion energy of \ee{51}erg.}
        \label{fig:pcrmax_\zetaval}
    \end{center}\end{figure*}

\begin{table*}
\caption{Partitions of the different energy components and the CR acceleration timescales for $3.6 < \xi <3.8$.}
\begin{center}
\begin{tabular}{|c|c|c|c|c|c|c|c|c|c|}
\hline
$\bism\;(\mu\mt{G})$  & $\nism\;(\mt{cm}^{-3})$ & $\Ekin$ (\ee{51}erg) & $\Eth$ (\ee{51}erg) & $\Ecr$ (\ee{51}erg) & $\tcr$ (\ee{3}yr) & $\tcr/t_\mt{s}$ \\ 
 \hline
1.0 & $0.01 - 0.1$ &  $0.17 - 0.23$ &  $0.14-0.36$ & $0.30-0.65$ & $2 - 8$ & $5-14$ \\ 
10  & $0.1 - 1.0$   &  $0.17 - 0.23$ &  $0.18-0.45$ & $0.25-0.60$ & $1 - 5$ & $4-13$ \\ 
\hline
\end{tabular}
\end{center}
\label{tab:ecr}
\end{table*}

\newcommand\cntplot[7]{
        \includegraphics*[trim=#1 #2 55 5,clip=true,width=#3\hsize]{\basedir/maps/z\zetaval_b#4/#5#6_t#7\lmcsnr_denxi}%
}

\section{Maps for the Galactic supernova remnants}
\label{sec:galactic}

\def\fcap{==}
\newcommand\bplot[2]{
	\cntplot{\marginfirst}{#1}{\wfourplotfirst}{#2}{\fvar}{}{430}%
	\cntplot{\margintrest}{#1}{\wfourplotrest}{#2}{\svar}{}{430}\hfill
	\cntplot{\margintrest}{#1}{\wfourplotrest}{#2}{\fvar}{}{1000}%
	\cntplot{\margintrest}{#1}{\wfourplotrest}{#2}{\svar}{}{1000}\\
}
\def\fvar{radius}
\def\svar{velocity}
\def\fplotlabel{the FS radius (in pc)}
\def\splotlabel{the FS velocity (in km/s)}
  \begin{figure*}\begin{center}%
       \bplot{39}{001}
        \bplot{5}{030}          
        \caption{Contour maps for  {\fplotlabel}  and {\splotlabel}  \if  \fcap  as functions of the acceleration efficiency $\xi$ and ambient medium density $\nism$ for SNRs at 430 yr (first and second columns) and 1000 yr (third and fourth columns). 
The values of the ambient magnetic field (from top to bottom) $\bism=1,\,30\,\mu$G. Shaded regions mark the allowed parameter spaces for Tycho and SN1006 SNRs. \else Idem as \rfig{fig:radius_velocity_\zetaval_denxi}. \fi}
        \label{fig:\fvar_\svar_\zetaval_denxi}
    \end{center}\end{figure*}

In what follows we show how the library can be used for the identification of SNRs properties.
Maps for the FS radii and velocities with the regions of the allowed parameter space (blue shaded areas) for Tycho and SN~1006 remnants are presented in \rfig{fig:radius_velocity_\zetaval_denxi}, where we show plots with the values ambient magnetic field of $1\,\mu$G (top row) and $30\,\mu$G (bottom row).  

The first and the second columns of \rfig{fig:radius_velocity_\zetaval_denxi} contain the models at the age of Tycho (430 year) with the forward shock radius (3.6 - 6.1 pc) derived from the angular measurements of \citet[][FS location of 251'']{warren05} and the distance measurements of \citet[][$d = 4\pm1$ kpc]{hayato10}. The FS velocity is estimated by \citet[][4600 - 4800 km s$^{-1}$]{hayato10} from the measurements of Si, S, and Ar emission lines broadening. The third and the forth columns are the maps for the models at the age of 1000 yr with the shaded region being the data of SN~1006 remnant. The FS radius is measured by \citet[][$R_\mt{FS}= 14.5'$]{cassam-chenai08} and the distance estimate is taken from \citet[][2.2 kpc]{winkler03}. The FS velocity of  $5000\pm400\,\mt{km\,s}^{-1}$ was derived by \citet{katsuda09, katsuda13} from Chandra proper motion measurements of the North East non-thermal dominated rim, assuming the distance of 2.2 kpc. 
Contour-maps with CR pressure ($\Pcr/P_\mt{tot}$) and compression ratio ($\rtot$) at the FS for the ages of 430 and 1000 yr are presented on \rfig{fig:Pc_Ptot_r_tot_\zetaval_denxi}. The shaded regions are the ranges of the CR pressure and the compression ratio estimated in \citet{kosenko10}. 

\def\fcap{!=}
\def\fvar{Pc_Ptot}
\def\svar{r_tot}
\def\fplotlabel{the CR pressure}
\def\splotlabel{the compression ratio.}
  \begin{figure*}\begin{center}%
       \bplot{39}{001}
        \bplot{5}{030}          
        \caption{Contour maps for  {\fplotlabel}  and {\splotlabel}  \if  \fcap  as functions of the acceleration efficiency $\xi$ and ambient medium density $\nism$ for SNRs at 430 yr (first and second columns) and 1000 yr (third and fourth columns). 
The values of the ambient magnetic field (from top to bottom) $\bism=1,\,30\,\mu$G. Shaded regions mark the allowed parameter spaces for Tycho and SN1006 SNRs. \else Idem as \rfig{fig:radius_velocity_\zetaval_denxi}. \fi}
        \label{fig:\fvar_\svar_\zetaval_denxi}
    \end{center}\end{figure*}

\def\lmcsnr{snr}
\def\fvar{radius_velocity}
\def\svar{Pc_Ptot_r_tot}
\def\fplotlabel{FS radius and velocity}
\def\splotlabel{CR pressure and compression ratio}

\newcommand\allplot[2]{
        \includegraphics*[trim={\marginfirst}  #1 40 5,clip=true,width=\wfourplotfirst\hsize]{\basedir/maps/z\zetaval_b#2/\fvar_Pc_Ptot_\svar_t430snr_denxi}%
        \includegraphics*[trim={\margintrest}  #1 40 5,clip=true,width=\wfourplotrest\hsize]{\basedir/maps/z\zetaval_b#2/\fvar_CD_\svar_t430snr_denxi}~
        \includegraphics*[trim={\margintrest}  #1 40 5,clip=true,width=\wfourplotrest\hsize]{\basedir/maps/z\zetaval_b#2/\fvar_Pc_Ptot_\svar_t1000snr_denxi}%
        \includegraphics*[trim={\margintrest}  #1 40 5,clip=true,width=\wfourplotrest\hsize]{\basedir/maps/z\zetaval_b#2/\fvar_CD_\svar_t1000snr_denxi}\\
}

\def\fvar{radius_velocity}
\def\svar{r_tot}
  \begin{figure*}
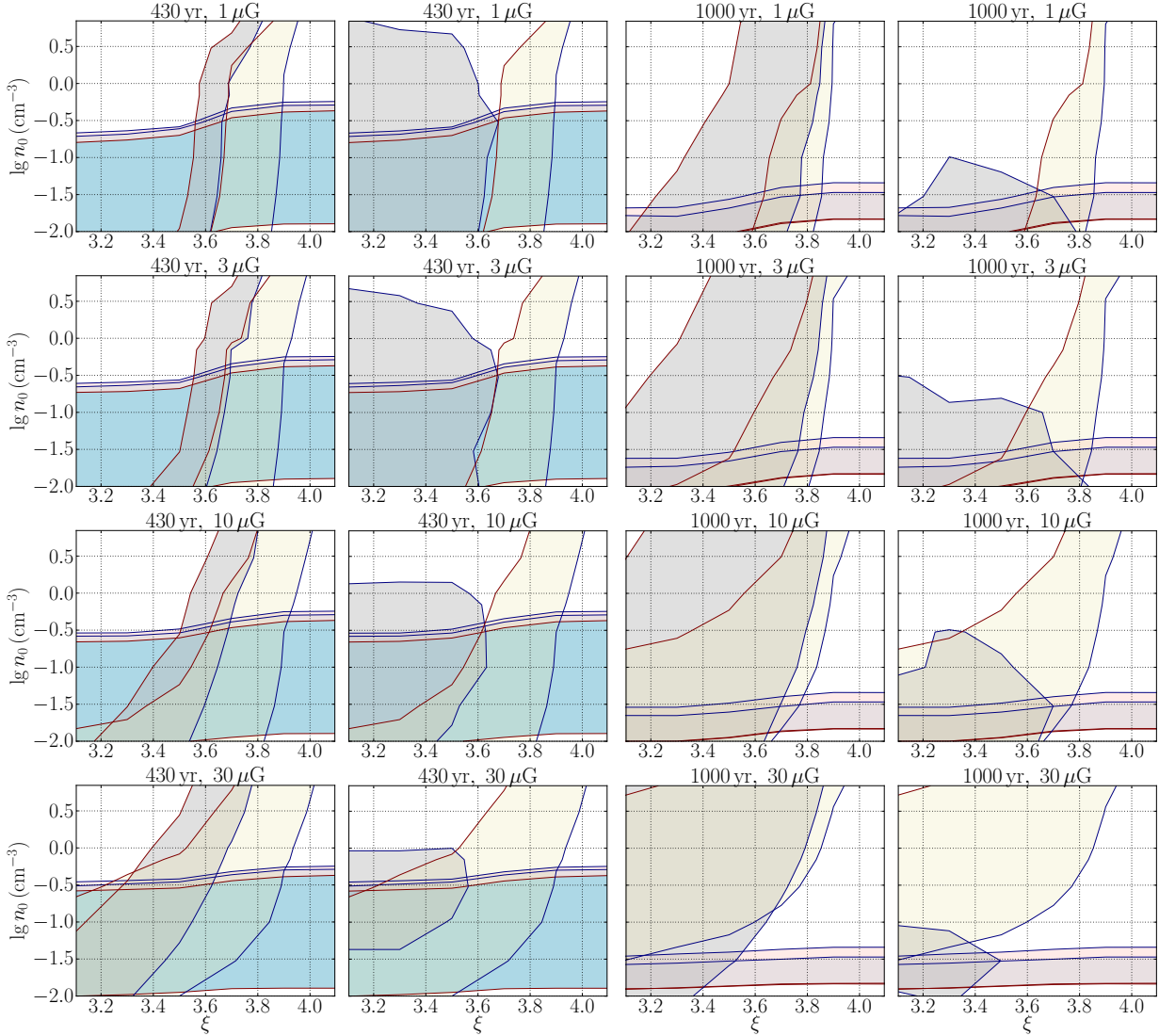
\begin{center}%
        \allplot{39}{001}
         \allplot{39}{003}
         \allplot{39}{010}
        \allplot{5}{030}
        \caption{Allowed parameter spaces for Tycho (the first and the second columns) and SN~1006 (the third and the fourth columns) SNRs. Maps with FS radius (blue), FS velocity (peach), CR pressure (grey, the first and the third columns), compression ratio (beige) and $R_\mt{CD}:R_\mt{FS}$ ratio (grey, the second and the fourth columns). For each plot, the intersection of all the shaded regions defines the constraints for the respective parameters for each remnant.} 
        \label{fig:all_\zetaval_cr_denxi}
    \end{center}\end{figure*}

These properties are assembled together in \rfig{fig:all_\zetaval_cr_denxi}. The FS radius (blue), FS velocity (peach), compression ratio (beige) and the CR pressure (grey) for Tycho are presented in the first column. The second column contains the same data with the CR pressure replaced by the $R_\mt{CD}:R_\mt{FS}$ ratio (CD: contact discontinuity) with the corresponding measurements for Tycho of \citet{warren05}.
The third and the forth columns show the data corresponding to SN~1006 with the $R_\mt{CD}:R_\mt{FS}$ ratio derived by \citet{miceli09}.
The intersection of the all regions constrain the ranges of the allowed parameter space ($\bism$, $\nism$, $\xi$), which are summarized in table~\ref{tab:galactic}. Note that the CR pressure $\Pcr/P_\mt{tot}$ and $R_\mt{CD}:R_\mt{FS}$ ratio select approximately the same respective regions in almost all the maps.

\begin{table*}
\caption{Properties of the Galactic SNRs ($\zeta = 0$).}
\begin{center}
\begin{tabular}{|c|c|c|c|c|c|c|c|c|}
\hline
 & \multicolumn{3}{|c|}{Tycho}& \multicolumn{3}{|c|}{SN~1006} \\
\hline
$\bism\;(\mu\mt{G}$) & $\nism\;(\mt{cm}^{-3})$ & $\xi$  & $\eta$ & $\nism\;(\mt{cm}^{-3})$ & $\xi$ & $\eta$ \\
 \hline
0.3 & - & - & - & $0.02 - 0.03$ & $3.6 - 3.7$ &  $10^{-4}$  \\
1.0 & - & - & - & $0.02 - 0.03$ & $3.6 - 3.7$ &  $(1 - 3)\times10^{-4}$  \\
3.0 & $0.30 - 0.40$ & $3.7$ & $10^{-4}$ & $0.01 - 0.03$ & $3.3 - 3.7$ & $10^{-4} - 10^{-3}$ \\
10 & $0.30 - 0.40$ &  $3.6 - 3.7$ & $(1-3)\times10^{-4}$ & $0.01 - 0.03$ & $<3.70$ & $>10^{-4}$ \\
30 & $0.25 - 0.40$ &  $3.3 - 3.6$ & $3\times10^{-4} - 10^{-3}$ & - & - & - \\
\hline
\end{tabular}
\end{center}
\label{tab:galactic}
\end{table*}

\def\zetaval{0}

For a more detailed study of the Galactic SNRs we performed simulations with higher AMR maximum mesh resolution of $256^3$.
Tycho models were zoomed out for $\bism=(3-10)\,\mu$G, $\nism=(0.3-0.4)\cc$, $\xi=(3.3-3.7)$. SN~1006 was modeled with $\bism =(0.3-10)\,\mu$G, $\nism=(0.01-0.03)\cc$, $\xi=(3.3-3.8)$. The corresponding contour-maps are presented in \rfig{fig:galmap_\zetaval}. The plots show that the simulations with the increased resolution produce almost the same results. However, there are minor modifications of the $R_\mt{CD}:R_\mt{FS}$ regions (e.g. 430 yr, $\bism=3\,\mu$G case) of the order of a few per cent. We note that the higher resolution models of $256^3$ cells consume a factor of 20 more computational resources compared to the resolution of $128^3$. 

\renewcommand\pcplot[6]{
        \includegraphics*[trim=#1 #2 47 0,clip=true,width=#3\hsize]{\basedir/\snrname/z\zetaval_b#4/radius_velocity_#5_r_tot_t#6snr_denxi}%
}

  \begin{figure*}
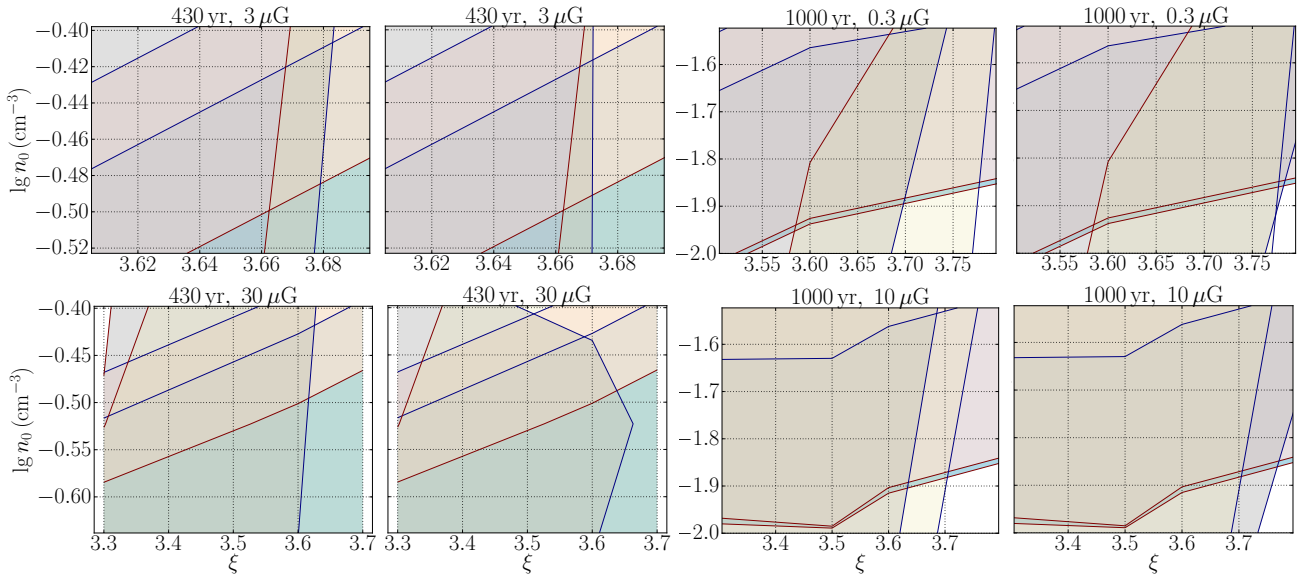
\begin{center}%
\def\snrname{Tycho}%
	\pcplot{\marginfirst}{37}{\wfourplotfirst}{003}{Pc_Ptot}{430}%
	\pcplot{\margintrest}{37}{\wfourplotrest}{003}{CD}{430}%
\def\snrname{1006}%
	\pcplot{\marginfirstlab}{37}{\wfourplotfirstlab}{000}{Pc_Ptot}{1000} %
	\pcplot{\margintrest}{37}{\wfourplotrest}{000}{CD}{1000} \\
\def\snrname{Tycho}%
	\pcplot{\marginfirst}{5}{\wfourplotfirst}{030}{Pc_Ptot}{430}%
	\pcplot{\margintrest}{5}{\wfourplotrest}{030}{CD}{430}%
\def\snrname{1006}%
	\pcplot{\marginfirstlab}{5}{\wfourplotfirstlab}{010}{Pc_Ptot}{1000}%
	\pcplot{\margintrest}{5}{\wfourplotrest}{010}{CD}{1000}
        \caption{Maps with FS radius (blue), FS velocity (peach), CR pressure (grey, the first and the third columns), compression ratio (beige) and $R_\mt{CD}:R_\mt{FS}$ ratio (grey, the second and the fourth columns). The first two columns are for Tycho, the third and the forth show the models for SN~1006. The data are based on the simulations of $256^3$ cells.  }
        \label{fig:galmap_\zetaval}
    \end{center}\end{figure*}



3D density slices of $512^3$ resolution runs for Tycho and SN~1006 are presented in \rfig{fig:denmap_\zetaval}.\footnote{The non-uniform structure of the FS results from a numerical instability \citep[discussed in][]{fraschetti10}.} For these detailed simulations we selected the configurations from the central parts of the allowed parameter regions. From top to bottom the models are $\bism=3\,\mu$G, $\nism=0.35\cc$, $\xi=3.67$ and $\bism=30\,\mu$G, $\nism=0.28\cc$, $\xi=3.3$ for Tycho (left column), $\bism=0.3\,\mu$G, $\nism=0.02\cc$, $\xi=3.7$ and $\bism=10\,\mu$G, $\nism=0.02\cc$, $\xi=3.3$ for SN~1006 (right column). Although the typical magnetic field in the Galaxy is about $5\,\mu$G, 
formally the dynamical and morphological SNR parameters alone cannot exclude the cases with $\bism$ as high as $30\,\mu$G. The next step in the fine-tuning of the SNR properties and specifically pin-pointing the magnetic field will be the detailed modeling of the thermal X-ray and non-thermal broad-band (including $\gamma$-ray) emission. This functionality is implemented in the code and described in details in \citet{ferrand12, ferrand14}.

\newcommand\denplot[6]{
        \includegraphics*[trim=#1 #2 40 30,clip=true,width=#3\hsize]{\basedir/\snrname/b#4_#5/slice_d_size09_00#6.png}%
}
  \begin{figure*}
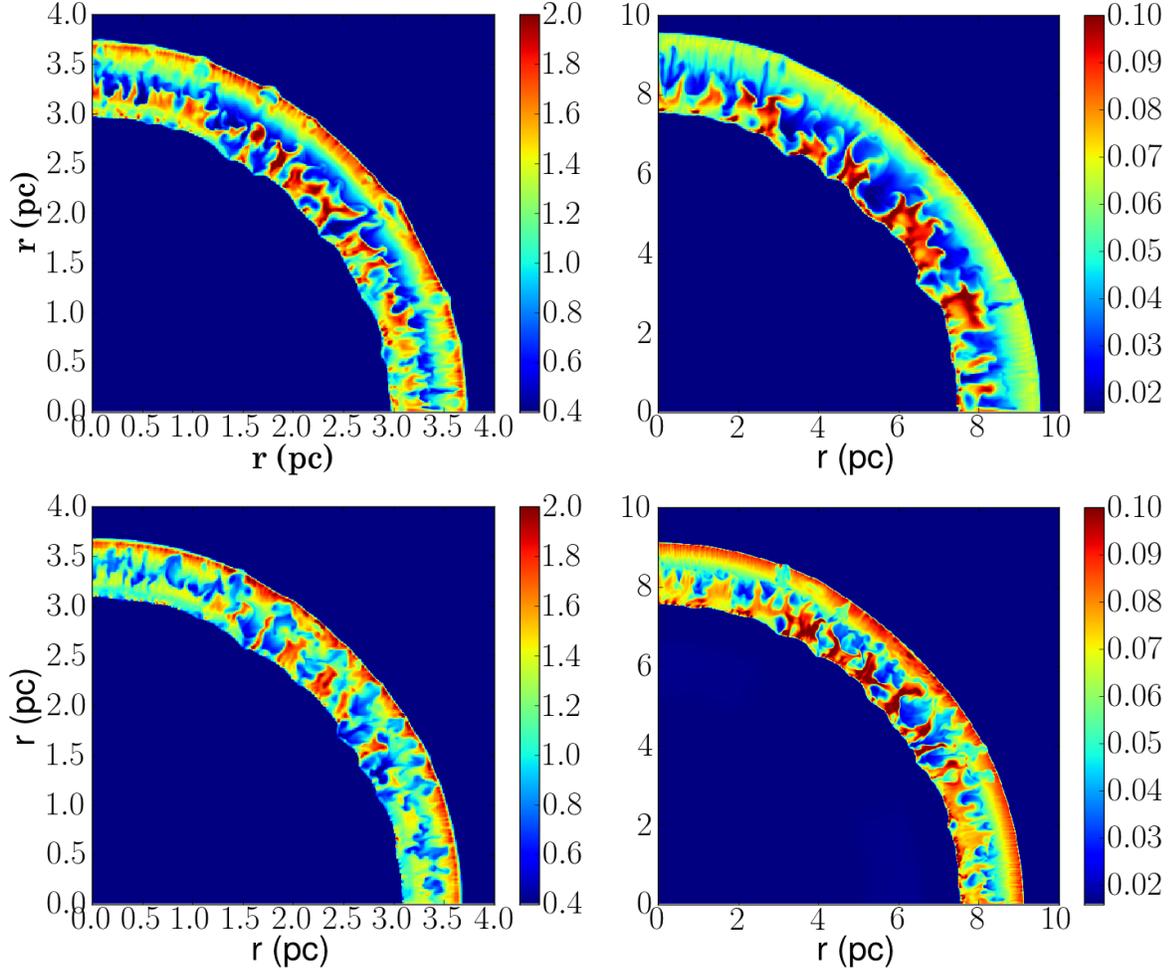
\begin{center}%
\def\snrname{Tycho}%
	\denplot{\marginfirst}{5}{\wtwoplotfirst}{003}{n035_xi367}{3}%
\def\snrname{1006}%
	\denplot{\marginslice}{5}{\wtwoplotrest}{000}{n002_xi370}{3}\\
\def\snrname{Tycho}%
	\denplot{\marginfirst}{5}{\wtwoplotfirst}{030}{n028_xi330}{3}%
\def\snrname{1006}%
	\denplot{\marginslice}{5}{\wtwoplotrest}{010}{n002_xi330}{3}\\
        \caption{Density slices for Tycho (left panel) with $\nism=0.35\cc$, $\xi=3.67$, $\bism=3$ (top), and $\nism=0.28\cc$, $\xi=3.3$, $30\,\mu$G (bottom). Right panels show slices for SN~1006 with $\nism=0.016\cc$, $\xi=3.7$, $\bism=0.3$ (top) and $\nism=0.016\cc$, $\xi=3.3$, $10\,\mu$G (bottom). Units of the colour-bars are $\cc$.}
        \label{fig:denmap_\zetaval}
    \end{center}\end{figure*}

\section{Supernova remnants in the Large Magellanic Cloud}
\label{sec:lmc}
\def\addlab{age}
\def\pvar{radius_velocity}

We also applied the library to the young SNRs located in the LMC: 0509-67.5 and 0519-69.0. Knowing  the distance to the LMC ($50$ kpc)  from the angular sizes we unequivocally derive physical dimensions of the remnants. We use the FS velocity measurements reported by \citet{ghavamian07}, estimated from the far-ultraviolet lines broadening at the FS, and the velocities of the plasma at the CD ($u_\mt{CD}$) derived by \citet{kosenko08, kosenko10} in the analysis  of the soft X-ray spectra in XMM-Newton RGS data.


\def\zetaval{00}

\def\fcap{==}
\def\addlab{0509-67.5}
\renewcommand\bplot[2]{
\def\lmcsnr{_0509}
	\cntplot{\marginfirst}{#1}{\wthreeplotfirst}{#2}{\pvar}{\crfl}{310} 
	\cntplot{\margintrest}{#1}{\wthreeplotrest}{#2}{\pvar}{\crfl}{360} 
	\cntplot{\margintrest}{#1}{\wthreeplotrest}{#2}{\pvar}{\crfl}{400}\\
}
\def\crfl{_ej}

\def\fcap{!=}
\def\crfl{_cr_ej}
\def\addlab{0509-67.5 (accounted for the DSA at the FS)}

\def\addlab{0519-69.0}
\renewcommand\bplot[2]{
\def\lmcsnr{_0519}
	\cntplot{\marginfirst}{#1}{\wthreeplotfirst}{#2}{\pvar}{\crfl}{580}
	\cntplot{\margintrest}{#1}{\wthreeplotrest}{#2}{\pvar}{\crfl}{650}
	\cntplot{\margintrest}{#1}{\wthreeplotrest}{#2}{\pvar}{\crfl}{720}\\
}
\def\crfl{_ej}

 \newcommand\lmcplot[2]{
 \def\lmcsnr{_0509}
	\cntplot{\marginfirst}{#1}{\wfourplotfirst}{#2}{\pvar}{\crfl}{310} 
	\cntplot{\margintrest}{#1}{\wfourplotrest}{#2}{\pvar}{\crfl}{400}~
\def\lmcsnr{_0519}
	\cntplot{\margintrest}{#1}{\wfourplotrest}{#2}{\pvar}{\crfl}{580}
	\cntplot{\margintrest}{#1}{\wfourplotrest}{#2}{\pvar}{\crfl}{720}\\
}
  \begin{figure*}
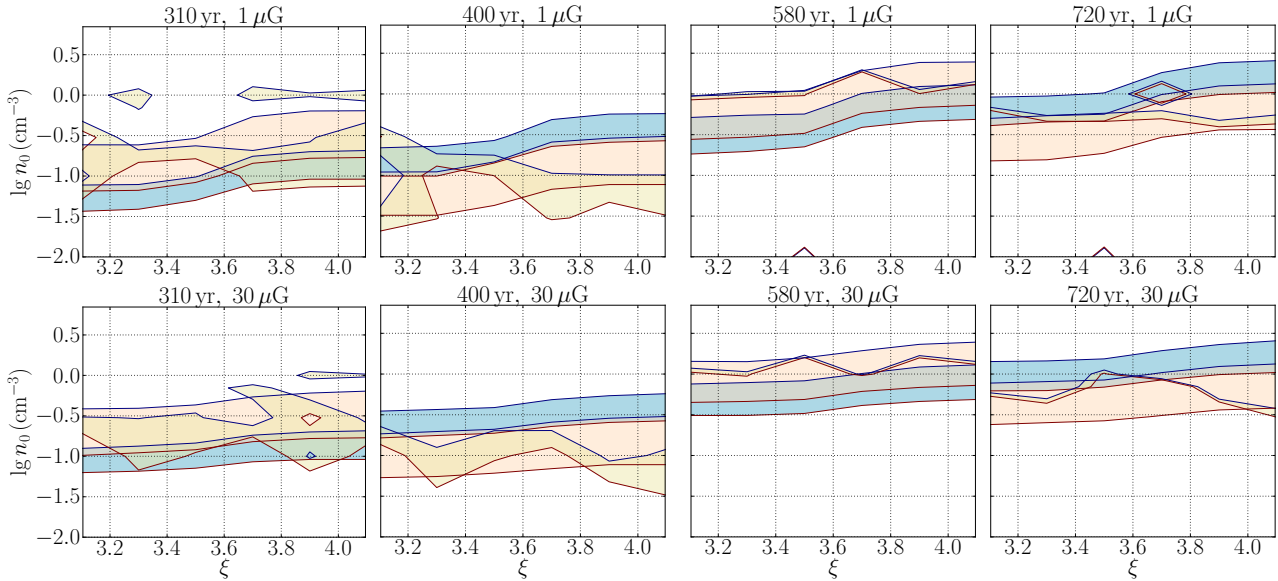
\begin{center}%
        \lmcplot{39}{001}
        \lmcplot{5}{030}
        \caption{Allowed parameter spaces as in \rfig{fig:all_\zetaval_cr_denxi} and \rfig{fig:galmap_0} for 0509-67.5 (first and second column) and  0519-69.0 (third and fourth columns) for $\bism=1\,\mu$G (top row) and $\bism=30\,\mu$G (bottom row). The \emph{blue} region outlines the FS radius,  \emph{beige} marks the ejecta velocity at the CD,  \emph{pink} colour shows the FS velocity derived from ultra-violet lines broadening.}
        \label{fig:\pvar_\zetaval_lmc\crfl}
    \end{center}\end{figure*}

      
As before, we built contour-maps of FS radii and velocities for the 0509-67.5 and 0519-69.0 SNRs. 
During the ejecta-dominated or early Sedov stages of the remnant's evolution, the main contribution to the emission line broadening is made by the expansion of the shocked material, rather than by the thermal motion of ions. Thus, we compare the measurements reported in \citet{kosenko08, kosenko10} with the maps of the dynamics of the ejecta in the vicinity of the CD. For each remnant, we selected the age limits when the measured FS radii and velocities are compatible. The corresponding contour-plots are presented in \rfig{fig:\pvar_\zetaval_lmc\crfl}.
This technique yields the age of the SNR~0509-67.5 model of $350\pm50$ yr with $\nism=(0.1-0.25)\cc$ and for the SNR~0519-69.0 of $650\pm70$ yr  ($630\pm90$ yr if the plasma velocity at the CD is not taken into account) with $\nism=(0.5-1.0)\cc$. 


\if !=
\begin{table}
\caption{Relation between $\bism$ and the ages of the LMC SNRs.}
\begin{center}
\begin{tabular}{|c|c|c|c|c|c|c|c|c|}
\hline
 & \multicolumn{3}{|c|}{0509-67.5}& \multicolumn{3}{|c|}{0519-69.0} \\
\hline
$\bism,\,\mu$G  & 310 & 360 & 400 & 580 & 650 &720 \\
 \hline
0.3 & + & + & + &  - & + & + \\
1.0 & + & + & + &  - & + & + \\
3.0 & -  & + & + & + & + &  - \\
10  & -  & + & + & + & + &  - \\
30  & -  &  - & + & + & + &  - \\
\hline
\end{tabular}
\end{center}
\label{tab:lmc}
\end{table}
\fi

\section{Discussion}
\label{sec:results}

The recent detailed modeling of \citet{bell13} showed that CR acceleration up to the knee in the Galactic spectrum is only possible if the magnetic field ahead of the shock is substantially amplified by the escaping CRs. As we described in \sect{sec:method}, in our simulations we used the recipe of \citet{caprioli09} where MFA is provided entirely by the CR-induced streaming instability. However, it is not entirely clear which mechanism provides the most of amplification. MFA is an important ingredient, yet its modeling is still much uncertain. Nevertheless, in this study we investigate only dynamical properties of the remnants, where the details of the magnetic field structure and amplification ahead of the shock do not play such a crucial role: all the unknown relevant parameters are secondary to the injection efficiency~$\xi$. These aspects become important in the modeling of the non-thermal emission from a SNR, which is presented in \citet{ferrand14} together with a more comprehensive discussion of MFA. For a study of the impact of different MFA models for a type Ia SNR we also refer the reader to the study of \citet{kang13a}. 

The approach used by \citet{caprioli09} is based on results valid for Alfv\'en wave dynamics in the case of a plane-parallel shock. Even though this may not be an exact description for a spherical shock, we assume that it is an acceptable approximation for the purpose of our parametric study. There could be some systematic bias here, however we believe that, when comparing with the observations, the uncertainties of the measurements are more important and we can apply here the parallel shock approach.

\subsection{Cosmic-ray energy evolution}
The maps with the CR energy (\rfig{fig:pcrmax_\zetaval}) and with the DSA timescales (\rfig{fig:tmax_\zetaval}) imply that a typical SNR during its lifespan can transfer up to $(30-60)\%$ of the explosion energy to the cosmic-ray particles. The acceleration of the relativistic particles falls-off after $t_\mt{CR} = 10^3-10^4$~years of the SNR expansion, long before the transition to the radiative phase $t_\mt{rad} \simeq 3.6$\e{4}$(E_0/10^{51})^{3/4}n_0^{-1/3}$ yr \citep{cioffi88}. 

According to \citet{li11} the supernova rate in our galaxy is $3$ per century. Hence, our findings on the channeling  of $(0.3-0.6)$\e{51}erg of the SNR energy to the relativistic particles result in $(3-6)\times$\ee{41}\ergs luminosity of the CRs in the Galaxy. This rough estimate is in general agreement with \citet{dogiel02} and \citet{strong10} calculations of the diffuse hard X-ray and $\gamma$-ray emission in the Galaxy, where they derive the CR luminosity of $\lesssim$\ee{41}erg s$^{-1}$. As suggested by \citet{cristofari13}, our value can be overestimated as it does not account for the parts of the shells interacting with dense clouds, where CR acceleration is suppressed. A specific geometry of the magnetic field, as in the case of SN~1006, also affects the acceleration efficiency. Note that we focus here only on type Ia supernovae, whereas a remnant of type II supernova expanding in a low density cavity created by the progenitor wind may be a less effective accelerator. Besides, the uncertainties of the SN rate and the CR luminosity estimates referenced above can reach a factor of two.

The energy profiles in \rfig{fig:crpar_gal_\zetaval_} differ from those presented by \citet{kang10, kang13}. In our models of the initially cold fast ejecta, expanding into the warm ISM  the kinetic energy is transferred into thermal and CR components, while in \citet{kang10,kang13} the thermal energy is already dominating in the very beginning of the SNR evolution. However in both cases at about several thousand years the CR energy saturates around some critical value. The level of this saturation depends on the assumed ISM properties \citep[warm or hot ISM in][]{kang10}. In our simulations the saturation occurs for the balanced cases with $\nism=0.03\cc$, $\bism=1\,\mu$G and $\nism=0.3\cc$, $\bism=10\,\mu$G. 

The range of the energy transferred to the CRs (table~\ref{tab:ecr}) covers the estimates of the previous studies \citep[e.g.][]{berezhko97,berezhko09,kang06} and is in agreement with values of \citet{kang13}, where it was found that without Alfv\'en drift $\Ecr$ reaches $30\%$ of the explosion energy and a few percent less if the drift is accounted for. In the efficient accelerating models of \citet{kang13} the fraction of the CR escape energy ranges in $0.2-6\%$, which corresponds in our maps to $\xi>3.6$ for $\nism>1.0\cc$ and $\xi>3.4$ for $\nism<0.3\cc$ models, as  the plots in the last column of \rfig{fig:pcrmax_\zetaval} show. 

\subsection{Implications for observations}
We demonstrated how the library can be used on the examples of four SNRs: Tycho, SN~1006, 0509-67.5 and 0509-69.0. Employing the available measured properties (such as FS velocity, the FS radius), the library allows to put constrains on the density of the ambient medium, the acceleration parameter $\xi$ and the ambient magnetic field.
We compared the observations with 3D hydrodynamical models, assuming a warm homogeneous ambient medium (consisting of only hydrogen atoms), a power-law ejecta density profile, and a supernova explosion energy of \ee{51}erg.

\subsubsection{Tycho}
The mapping of the Tycho's properties suggests an ambient magnetic field of $(3-10)\,\mu$G, 
density of $(0.3-0.4)\cc$,  the acceleration parameter $\xi = (3.6 - 3.7)$, injection fraction $\eta=(1-3)\times10^{-4}$, compression ratio $\rtot=(6-7)$, with an assumed distance of $(3.0-3.1)$~kpc. The energy components are $\Ekin = (0.53-0.56)$, $\Eth = (0.20-0.25) , (0.12-0.22)$ , $\Ecr = (0.18-0.23) , (0.20-0.30)$ (in units of \ee{51}erg) for $\bism = 3 , 30 \,\mu$G.
The effects of the turbulent wave heating were studied for the high-resolution simulations with the maximum number of cells of $512^3$ in the model with a magnetic field as high as $30\,\mu$G. In this case the dynamical parameters and energy distribution change only very slightly. For the model with $\nism=0.28\cc$ and extreme value for the acceleration parameter $\xi=3.3$ we have found $\Ekin=0.53,\;\Eth = 0.14/0.12,\;\Ecr=0.27/0.29$ for $\zeta=0/1$. 

The inferred ISM density of $0.3\cc$ is higher that the upper limit $\nism\lesssim 0.2\cc$, estimated by \citet{katsuda10} and considerably lower than the best matched value of $\nism = 0.85\cc$ derived from hydrodynamical and X-ray spectral modeling of \citet{badenes06} with an adopted explosion energy of $1.2$\e{51}~erg. Nevertheless, there are indications of the non-trivial history of this remnant and it is possible that in the earlier stages the shell was interacting with the pre-supernova dense wind, created by the progenitor system \citep[e.g.][]{chiotellis13}. In this case the homogeneous ISM models should be used with caution.

\citet{volk08} employed the nonlinear kinetic theory of cosmic ray acceleration (1D calculations with a correction of 1.05 for Rayleigh-Taylor instabilities), adopted an explosion of $1.2$\e{51}erg (MFA was not included in the theory) and found a model matching the mean ratio between the radii of the CD and the FS of the remnant and the observed $\gamma$-ray emission with $\eta=3\times10^{-4}$, $\nism \lesssim 0.4\cc$, compression ratio $\rtot = 5.15$, and the distance $d=3.3$~kpc. 
With the SNR library we found similar injection fraction of $(1-3)\times10^{-4}$. However, comparison of their compression ratio with the contour-maps in the second column of \rfig{fig:all_00_cr_denxi} (the corresponding scales are provided in  \rfig{fig:Pc_Ptot_r_tot_00_denxi}) shows no intersection of the $\rtot = 5.15$ values with the allowed $R_\mt{CD}:R_\mt{FS}$ region for any $\bism$ and $\xi$. This implies that, in our framework, no hydrodynamical model can accommodate both these values simultaneously. The CD location in the 3D models with the account for the Rayleigh-Taylor instability sets the lower limits of $\sim 6$ on the total compression ratio in the models for Tycho. 

Our findings support the assumption made in the study of \citet{morlino12}, where they applied $\nism = 0.3\cc$, $\xi=3.7$ and  $d=3.3$ kpc, and considered a \ee{51}erg explosion with $1M_\odot$ mass of ejecta. They modeled the $\gamma$-ray emission detected with the Fermi-LAT and VERITAS observatories employing a semi-analytical approach to the NLDSA accounting for MFA due to resonant streaming instability and the back-reaction on the shock from both the CR and the self-generated magnetic turbulence. It was found that at the present epoch the remnant channels 10\% of its kinetic energy into CRs. Further, comparison of the $R_\mt{CD}:R_\mt{RS}$ measurements (RS: reverse shock) with spherically-symmetrical 1D hydrodynamical models obtained with a two-fluid treatment for CRs and assuming a 1.4\e{51} erg explosion \citep{kosenko11} results in $\Ecr/\Ekin=(10-20)\%$ at the FS for Tycho. These ranges are somewhat lower than our findings $\Ecr/\Ekin = (30-40)\%$ derived from the same $R_\mt{CD}:R_\mt{FS}$ measurements.  

In the latter case the explosion energy plays an important role. Assuming a more energetic model would imply a higher ISM density and a higher kinetic energy with the absolute CR energy deposited unchanged (as DSA depends strongly on the FS Mach number), thus reducing the acceleration efficiency estimate. 
On the other hand, \citet{morlino12} used \ee{51}erg models to fit the emission and this argument does not hold, although their approach of modeling $\gamma$-ray spectra was contested by \citet{berezhko13}. 
Note that the low resolution models (e.g. the fourth column of \rfig{fig:crpar_gal_\zetaval_} with $\xi=3.7$) at the age of Tycho produce more conservative values of $\Ecr \simeq 0.15\times$\ee{51}erg and $\Ecr/\Ekin \simeq 0.3$. 
This can imply that the 3D instabilities in the shell affect the final results. 

Recently, \citet{slane14} performed a 1D spherically symmetric modeling of Tycho. They adopted $E_0= 10^{51}$erg, and found a best fit for $\nism \sim 0.3\cc$, with $\Ecr/\Ekin \simeq 0.16$. They found a downstream magnetic field of $180\,\mu$G, with an unperturbed ISM value of $5\,\mu$G.
In their modeling \citet{morlino12} found a best-fitting downstream magnetic field of $300\,\mu$G, with the same unperturbed ISM value of $5\,\mu$G. 
\citet{volk08} determined the downstream magnetic field value from the observed linear thickness of X-ray filaments of $412\,\mu$G. 
The mapping of Tycho's dynamical properties reported here suggests the values of $200,\,900/100\,\mu$G for $\bism=3, 30\,\mu$G correspondingly and for $\zeta=0/1$ in the latter case. In our grid the measured downstream magnetic field of $\sim400\,\mu$G can be associated with the cases of $\bism\simeq(10-30)\,\mu$G and possibly with effective turbulent heating.

\subsubsection{SN~1006}
The mapping of SN~1006 properties yields an ambient medium density  of $(0.01-0.03)\cc$, and an acceleration parameter of $3.6-3.7$. The energy distribution in the remnant is as follows (in units of \ee{51}erg): $\Ekin = (0.55-0.64)$, $\Eth = (0.22-0.31),(<0.25)$, $\Ecr = (0.10-0.17),(0.10-0.25)$ for $\bism=0.3,10\,\mu$G. 

The allowed range found for the ISM density is lower than the results of \citet{berezhko09} derived from the TeV \citep[H.E.S.S.,][]{acero10} flux $0.035 \lesssim\nism\lesssim 0.05\cc$, and \citet{acero07} value, based on the X-ray emission from the South East rim of the remnant, of $0.05\cc$ assuming a supernova explosion energy of $1.7$\e{51} and $2$\e{51}erg respectively. Moreover, compression ratios of 4.9 and 4.7 for $\nism=0.05$ and $0.035\cc$ \citep{berezhko09} are somewhat lower than our findings of $(5-8)$ restricted by the $R_\mt{CD}:R_\mt{FS}$ measurements (the rightmost column of \rfig{fig:all_\zetaval_cr_denxi}). 
Yet the spatially resolved spectral analysis of XMM-Newton spectra by \citet{miceli12} implies a compression ratio of $\sim6$ in the regions of the strongest non-thermal emission. 
Note that basing the hydrodynamical models on a 2\e{51}erg explosion would increase the $\nism$ estimate by a factor of two, changing it to $(0.02-0.06)\cc$, which is still lower than the value of $0.085\cc$ of \citet{katsuda09}. 

In the revision of the TeV emission analysis with non-linear kinetic theory, \citet{berezhko12} updated the circumstellar hydrogen density to $0.06\cc$ with a share of the CR energy of $5\%$ of the explosion energy (of $2$\e{51}erg). The latter is in agreement with our estimate $\Ecr = (0.10-0.17)$\e{51}erg.  
Their best fit injection parameter $\eta=2$\e{-4} is close enough to our estimate of $1.3$\e{-4} for $\bism=0.3\,\mu$G. 
In the fitting of the spatially-integrated synchrotron emission, \citet{berezhko12} found a downstream magnetic field of $150\,\mu$G. In our high-resolution models for SN~1006 with $\nism = 0.02\cc$ the amplified magnetic field behind the FS is $20, 180/38\,\mu$G for $\bism=0.3, 10\,\mu$G and $\zeta=0/1$. This comparison suggests that the ambient magnetic field around SN~1006 can reach $10\,\mu$G without Alfv\'en wave damping.

However, in the detailed analysis of the X-ray radial profile reported by \citet{morlino10}, they fixed $\nism = 0.05\cc$, ($\bism=3\,\mu$G,  downstream magnetic field $90\,\mu$G) and found the best fit to the observed radial profile with $\xi = 3.95$ and the CR acceleration efficiency of $\sim29\%$.  We have found the lower limit $\xi=3.7$ for $\bism=10\,\mu$G case and respective $\Ecr/\Ekin = (0.16-0.45)$. On the other hand, \citet{petruk11} in their analysis of the radio, hard X-ray, and $\gamma$-ray data also found $\bism = 12\,\mu$G and the magnetic field inside the remnant of $32\,\mu$G, which is in agreement with ``leptonic'' model parameters of \citet{acero10}. According to our models, such a configuration corresponds to the efficient turbulent heating acting in the precursor. 

\subsubsection{LMC SNRs}
Although SNR~0509-67.5 was examined by many authors, its physical properties remain much uncertain. Studies of the light echo by \citet{rest05} derived the age limits of $400\pm120$ yr, 
 thermal X-ray modeling by \citet{badenes08} implies $\sim$400 yr, and the analysis of \citet{ghavamian07} yields $295-585$ yr. 
The determination of the ISM density is also somewhat problematic. In the spectral fitting of the Chandra data with the non-thermal dominated X-ray continuum model, \citet{warren04} found $\nism\lesssim 0.05\cc$. The best matched ambient density in the hydrodynamical simulations coupled with X-ray spectral modeling of \citet{badenes08} was found to be $0.4\cc$. The detailed analysis of the XMM-Newton spectra reported by \citet{kosenko08} revealed $\nism = (0.4-0.8)\cc$ and they used $\nism \simeq 0.1\cc$ for the hydrodynamical models.

The range of $(0.1-0.3)\cc$ mapped in the SNR model grid presented here can reconcile these estimates. The age of $360\pm50$ yr is in agreement with the previous studies. The explanation of the discrepancy between the X-ray density values and the mapped dynamical range can be twofold.
On one hand, higher $\nism$ derived from X-ray spectral modeling could be of the same origin as for Tycho, where the X-ray spectra suggest that the ISM density is almost 5 times higher than the dynamical value \citep[the detailed case study is reported in][]{chiotellis13}.
On the other hand, our library is built with the assumption of an explosion energy of \ee{51}erg. A~more energetic supernova would imply a higher ISM density for the measured FS radius and velocity. 

SNR 0519-69.0 is also one of the remnants for which the age of $600 \pm 200$ years was derived from the light echo analysis \citep{rest05}. The detailed multicomponent analysis of the XMM-Newton X-ray spectra and Chandra X-ray image provided the ISM density of  $2.4 \pm 0.2\cc$ \citep{kosenko10}.

Using only the FS radius and velocity measurements, the library of dynamical models restricts the age for this SNR to $630\pm90$ yr. Accounting for the plasma velocity at the CD refines this number to $650\pm70$ yr. The allowed parameters space yields an ISM density of $(0.5-1.0)\cc$. The difference is similar to the case of SNR~0509-67.5. A twice as more energetic supernova than assumed here would imply $\nism = (1.0-2.0)\cc$. Moreover, the typical overestimation of the density in the X-ray emission analysis over the dynamical value could again be similar to the case of Tycho \citep{chiotellis13}.

\section{Conclusion}
\label{sec:conclusion}

In this paper we presented a library of 3D models of type Ia SNRs evolving through a homogeneous medium. The multi-dimensional approach in the hydrodynamical simulations allows to accurately estimate the content of the different energy components in the remnants. Moreover, the detailed 3D modeling yields the location of the CD with the proper account for the Rayleigh-Taylor instability \citep{fraschetti10, ferrand10}. The library is built on a grid of the crucial parameters for the SNR evolution: ambient medium density $\nism$, ambient magnetic field $\bism$, and acceleration parameter $\xi$. We considered rather wide ranges of these parameters, so that the grid can accommodate all the SNRs with sufficiently known properties. The library enables us to evaluate the efficiency of the CR acceleration in SNRs if environmental parameters are sufficiently well known, and to pin down these parameters if the CR efficiency could otherwise be estimated.

Using the measurement of the FS radius and velocity of Tycho and SN~1006 we localized these remnants on the ($\nism, \bism, \xi$)-grid and derived the respective allowed parameter spaces.  For SNR 0509-67.5 and SNR 0519-69.0 we identified their dynamical ages and $\nism$. In the future this method can be used as a guide for the interpretation of observations of other SNRs.


\vskip 1cm
\textit{Acknowledgments.}  
DK was funded by the French National Research Agency (ANR) under the COSMIS project and supported by Russian Foundation for Basic Research (RFBR) grant 13-02-92119. 
GF has been funded by an NSERC's Canada Research Chairs award to Samar Safi-Harb, with additional support from the Canada Foundation for Innovation and the Manitoba Research and Innovation Funds. GF had been also partially supported by the Canadian Institute of Theoretical
Astrophysics (CITA) through the National Fellowship Program.
All simulations were performed on the IRFU/COAST cluster at CEA Saclay.
The authors would like to thank Samar Safi-Harb for proofreading the final manuscript and anonymous referee for the revision and valuable comments. 


\bibliographystyle{mn2e}
\bibliography{mainbib}

\end{document}